\documentclass[12pt]{article}
\pdfoutput=1

\usepackage{epsfig,latexsym,amsfonts,amsmath,amsthm,amssymb,amsbsy,multirow,slashed,color,mathrsfs,wasysym,textcomp,subfigure,wrapfig,comment}

\usepackage{hyperref}
\hypersetup{colorlinks,
 linkcolor = blue,
 anchorcolor = red,
 citecolor = blue,
 filecolor = red,
 urlcolor = red,
 linktocpage}
 
\newcommand{\be}{\begin{equation}}
\newcommand{\ee}{\end{equation}}
\newcommand{\bea}{\begin{eqnarray}}
\newcommand{\eea}{\end{eqnarray}}

\DeclareSymbolFont{AMSb}{U}{msb}{m}{n}
\DeclareMathSymbol{\N}{\mathbin}{AMSb}{"4E}
\DeclareMathSymbol{\Z}{\mathbin}{AMSb}{"5A}
\DeclareMathSymbol{\R}{\mathbin}{AMSb}{"52}
\DeclareMathSymbol{\Q}{\mathbin}{AMSb}{"51}
\DeclareMathSymbol{\p}{\mathbin}{AMSb}{"50}
\DeclareMathSymbol{\I}{\mathbin}{AMSb}{"49}
\DeclareMathSymbol{\C}{\mathbin}{AMSb}{"43}
\DeclareMathSymbol{\D}{\mathord}{symbols}{"72}
\DeclareMathSymbol{\HH}{\mathbin}{AMSb}{"48}

\begin{document}


\begin{titlepage}
\bigskip
\bigskip\bigskip\bigskip
\centerline{\Large \bf Journey to the Center of the Fuzzball}
\bigskip\bigskip\bigskip
\bigskip\bigskip\bigskip

 \centerline{{\bf Fang Chen,}\footnote{\tt fchen@kitp.ucsb.edu}${}^\dagger$ {\bf Ben Michel,\footnote{\tt michel@physics.ucsb.edu}*
 {\bf Joseph Polchinski,}\footnote{\tt joep@kitp.ucsb.edu }*${}^\dagger$ and {\bf Andrea Puhm}}\footnote{\tt puhma@physics.ucsb.edu}*}
 
 \bigskip
\centerline{\em *Department of Physics}
\centerline{\em University of California}
\centerline{\em Santa Barbara, CA 93106 USA}
\bigskip
\centerline{\em ${}^\dagger$Kavli Institute for Theoretical Physics}
\centerline{\em University of California}
\centerline{\em Santa Barbara, CA 93106 USA}

\bigskip\bigskip\bigskip

\begin{abstract}
We study two-charge fuzzball geometries, with attention to the use of the proper duality frame.  For zero angular momentum there is an onion-like structure, and the smooth D1-D5 geometries are not valid for typical states.  Rather, they are best approximated by geometries with stringy sources, or by a free \mbox{CFT}.  For non-zero angular momentum we find a regime where smooth fuzzball solutions are the correct description.  Our analysis rests on the comparison of three radii: the typical fuzzball radius, the entropy radius determined by the microscopic theory, and the breakdown radius where the curvature becomes large.  We attempt to draw more general lessons.

\end{abstract}
\end{titlepage}

\baselineskip = 16pt
\tableofcontents

\baselineskip = 16pt

\setcounter{footnote}{0}

\section{Introduction}

The conflict between quantum theory and general relativity exposed by the black hole information paradox has swung back and forth for nearly four decades, recently inflamed by the firewall paradox. 
There have been a variety of previous proposals that the black hole horizon is not as general relativity describes.  In particular, the fuzzball program argues that the structure of the horizon is necessarily modified by the extended objects of string theory.  Indeed, key features of the firewall argument were first put forward as evidence for fuzzballs~\cite{Mathur:2009hf}.

In this paper we focus primarily on the simplest version of fuzzballs, the two-charge system of D1-D5 branes compactified on a circle.  In \S~\ref{sec:Jeq0} we reexamine the argument that the naive two-charge geometry is unphysical, and that fuzzball solutions are the correct description.
We begin by noting that as one approaches the singularity of the naive geometry, the first sign of a breakdown is that the radius of a circle drops below the string scale.  This suggests a $T$-duality from the original IIB picture to IIA, and indeed this provides a description valid down to smaller radii.  Eventually the coupling grows large, and an $S$-duality takes us to M theory.  In this regime the four-torus shrinks toward zero size, and a further $STS$ duality brings us to a new Type II description, in which the charges are carried by fundamental strings and momentum.  Finally this breaks down due to the spacetime curvature  becoming large, and no further stringy duality can save us.  Rather, the final picture is a weakly coupled CFT.

This onion-like layered structure has already been described in detail by Martinec and Sahakian~\cite{Martinec:1999sa}, building on the classic analysis of non-conformal branes in Ref.~\cite{Itzhaki:1998dd}.   However, its significance for the fuzzball program does not seem to have been discussed.  

Fuzzball solutions
approximate the naive geometry outside some crossover radius, which depends inversely on the average harmonic excited, $\overline{m}$.  For different values of $\overline{m}$, the crossover radius may lie within any of the IIB/IIA/M/II$'$/CFT regimes, and the parametrically valid description is a fuzzball solution in the given duality frame.  For {\it typical} states, the crossover occurs right at the transition between the final geometric picture, II$'$, and the free \mbox{CFT}.  In particular, this changes the standard picture of two-charge D1-D5 fuzzballs.  The smooth geometries~\cite{Lunin:2001jy} are not an accurate description for typical states. Rather, the best (though still marginal) supergravity description is one with explicit stringy sources.

Indeed, it is well-known that typical two-charge fuzzballs lie right at the breakdown of supergravity.  In fact, there are three important radii that are known to coincide: the typical {\it fuzzball radius} $r_{\rm f}$; the {\it entropy radius} $r_{\rm S}$, where the area in Planck units just matches the density of states of the system; and the {\it breakdown radius} $r_{\rm b}$, beyond which supergravity cannot be continued.  Historically the D1-D5 fuzzballs were derived by a duality chain from F1-$p$ solutions.  These 
are the same as the charges of our II$'$ description.  We trace the relation between these descriptions, and we emphasize the distinction between two free orbifold CFTs that arise in the D1-D5 system.

Much of the discussion of two-charge fuzzballs focuses on this final transition radius, and compares  fuzzballs with a black hole solution including $\alpha'$ corrections.   Our focus is rather on descriptions that are parametrically valid.  In search of a more interesting situation, we consider in \S~\ref{sec:Jneq0} states with large angular momentum $J$, for which the naive geometry is a black ring.  This geometry breaks down due to large curvature as we approach the ring.  We find that, as measured from the ring, the fuzzball and entropy radii again coincide, but the breakdown radius can be larger or smaller, depending on parameters.  Thus we identify a regime where the fuzzball description is parametrically valid and physically correct, even though the naive geometry still has small curvature.  We suggest that the breakdown of the naive geometry is instead signaled  by the entropy radius, beyond which the naive geometry would describe more states than holography allows.
In \S~\ref{sec:discussion} we discuss further directions.

\section{The $J=0$ system}\label{sec:Jeq0}

\subsection{Naive geometry: small black hole}\label{subsec:naivegeo}

Consider the background
\bea
ds^2_{\rm IIB} &=& \frac{1}{\sqrt{H_1 H_5}}(-dt^2+R^2dy^2) + \sqrt{H_1 H_5} \, dx_4^2  + \sqrt{\frac{H_1}{H_5}}\sqrt{V} dz_4^2 \,,\nonumber\\
e^{\Phi_{\rm IIB}} &=& g \sqrt{\frac{H_1}{H_5}} \,, \nonumber\\
C_2 &=& g^{-1}\left[H_1^{-1} dt\wedge R dy + Q_5 R \cos^2\tilde \theta d\psi \wedge d\phi\right] \,,
\eea
where 
\bea
H_{1} &=& 1+\frac{g N_1}{V r^2} \equiv 1+\frac{Q_1}{r^2} \,,\nonumber\\
 H_5&=& 1+\frac{g N_5}{r^2} \equiv 1+ \frac{Q_5}{r^2}\,.
\label{naivemet}
 \eea
We work in units such that $\alpha'=1$.  The four flat transverse directions $x$ are non-compact, and can be coordinatized as $dx_4^2=d r^2+r^2(d\tilde \theta^2+\sin^2\tilde \theta d\phi^2+\cos^2\tilde \theta d\psi^2)$, where the tildes are included to conform with standard notation~\cite{Mathur:2005zp}.  The $T^4$ coordinates 
$z$ have period $2\pi$. We consider the case where the $T^4$ is replaced by K3 in Appendix~\ref{app:K3}.

For non-compact $y$, the infrared geometry is $AdS_3 \times S^3 \times T^4$ in Poincar\'e coordinates.   If we then identify $y \cong y + 2\pi$, the horizon $r=0$ is a fixed point and becomes a cusp singularity.  For the compact theory there are three moduli: the coupling $g$, the circle radius $R$, and the torus volume $V$.  In the attractor limit where we ignore the 1's in the harmonic functions, only the modulus $g$ remains.  The torus volume flows to the attractor value $V = N_1/N_5$, while $R$ appears only in the combinations $Rr$ and $y/R$.  For simplicity we fix $V$ to its attractor value, so that $Q_1 = Q_5 \equiv Q$ and $H_1 = H_5 \equiv H$.  We are most interested in the attractor region, but it is useful to keep the harmonic function $H$ general.  The background is then given by \eqref{naivemet} with $e^{\Phi_{IIB}}=g$ and $H=1+Q/r^2$ where the 1 drops out in the attractor.

In order for this D1-D5 description to be the correct duality frame asymptotically, we need the coupling and curvature to be small, and the circle and torus to be larger than the string scale.  Thus, 
\be
g < 1\,,\quad  Q>1\,,\quad R > 1\,,\quad N_1 > N_5 \,. \label{regime}
\ee
Discussions of this system often begin with a dual F1-$p$ description.  In \S~\ref{subsec:fuzzgeos} we will discuss connections with this frame.

\subsection{Into the black onion}\label{subsec:intoonion}

In the fuzzball program, it is argued that for $y$ compact the geometry~(\ref{naivemet}) breaks down even before the singularity, and must be replaced by fuzzball solutions.  We wish to ask, is there some signal of this breakdown as we approach the singularity?

While this work was in progress, we learned that this question had already been addressed by Martinec and Sasakian~\cite{Martinec:1999sa}.  Since this result does not seem to be widely known, we review their analysis.

\begin{figure}[!ht]
\begin{center}
\vspace {-5pt}
\includegraphics[width=\textwidth]{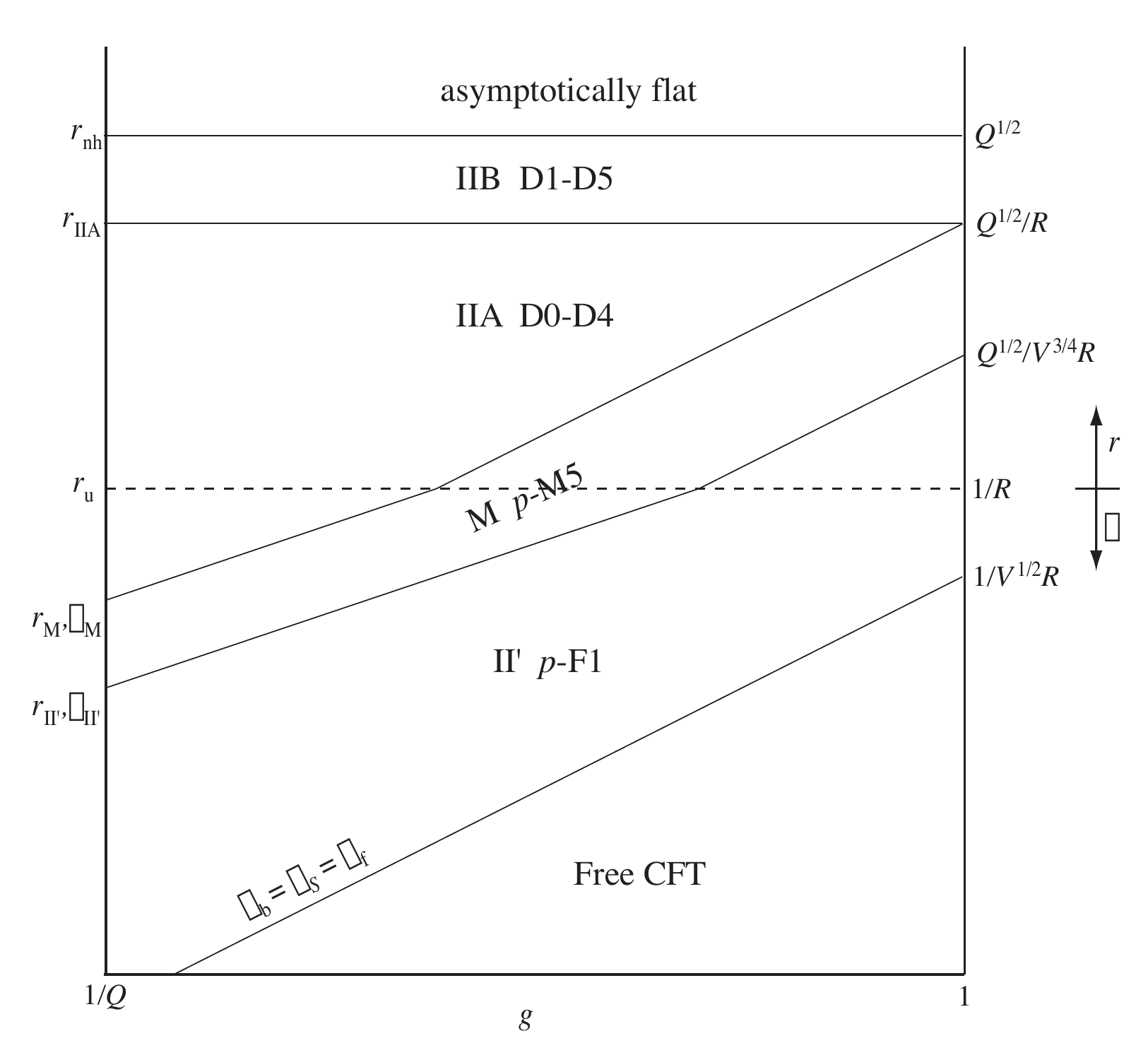}
\end{center}
\vspace {-10pt}
\caption{Domains of duality frames, on a log-log plot of radius and coupling.  The dashed line divides smeared geometries (above) from unsmeared (below).
}
\label{fig:radii}
\end{figure}

Note that the $y$ circle is shrinking, and at a radius $r \sim r_{\rm IIA}= {Q}^{1/2}/R$ it reaches the string scale.  In the D1-D5 regime~(\ref{regime}) this is always inside the crossover to the near-horizon region, $r \sim r_{\rm nh} = {Q}^{1/2}$.\footnote{We will encounter a long list of significant radii as we move along. Figure~\ref{fig:radii} gives an overview.  Because of the scaling in the attractor region noted above, most radii are proportional to $1/R$.}
This breakdown suggests a $T$-duality along the $y$ circle to a IIA solution, and indeed this extends the range of validity to smaller $r$.
The solution is
\bea
ds^2_{\rm IIA} &=& -H^{-1} dt^2 + H (d\tilde y^2/R^2+  dx_4^2)+\sqrt{V} dz_4^2\,, \nonumber\\
e^{\Phi_{\rm IIA}} &=& \frac{g\sqrt H}{R} \,,\nonumber\\
C_1 &=& \frac{R}{g H} dt \,,\nonumber\\
C_3 &=& g^{-1} Q R \cos^2\tilde\theta d\psi \wedge d\phi\wedge d\tilde 
y \,.
\eea

In the IIA frame, the charges are carried by D0- and D4-branes localized in the $\tilde y$ direction.  We are interested in single-particle states, so the branes should be coincident in the $\tilde y$ direction.  Unsmearing the sources gives
\be
H = \frac{Q}{r^2}\rightarrow \frac{\pi Q}{R}\sum_n \frac{1}{[r^2+(\tilde y-2\pi n)^2 /R^2]^{3/2}} \sim  \frac{\pi Q}{R \rho^3 } \,,
\ee
where the normalization is fixed by the large-$r$ behavior.  The crossover to the unsmeared solution is at $r \sim r_{\rm u} = 1/R$.  In the last line we have given the form as we approach the $\tilde y = 0$ image, where $\rho^2 = r^2 + \tilde y^2/R^2$.

As we continue toward the singularity, the IIA coupling becomes large, suggesting a lift to M theory.  If we work with the smeared metric, this occurs at $r \sim r_{\rm M} = g Q^{1/2}/R$.  Thus $r_{\rm M}/r_{\rm u} = g Q^{1/2}$.  In the D1-D5 regime~(\ref{regime}), $g$ is small and $ Q$ is large, but the product $g Q^{1/2}$ is not restricted.  If $g Q^{1/2}>1$, the transition to the M theory picture occurs in the smeared regime, at $r \sim r_{\rm M}$.  If $g Q^{1/2}<1$ it occurs at in the unsmeared regime, at $\rho = \rho_{\rm M} = g^{2/3} Q^{1/3}/R$.  

Either way, we end up with the M theory solution
\bea
ds^2_{\rm M} &=& e^{-2\Phi_{\rm IIA}/3} ds^2_{\rm IIA} + e^{4\Phi_{\rm IIA}/3}(dx_{11}+ C_1)^2 \nonumber\\
&=& \left(\frac{R^2}{g^2 H}\right)^{1/3}\left[-H^{-1} dt^2 + H(d\tilde y^2/R^2+dx_4^2)+\sqrt{V}dz_4^2\right] + \left(\frac{g^2 H}{R^2}\right)^{2/3}\left( dx_{11} + \frac{R}{gH}dt\right)^2 \,,
\nonumber\\[3pt]
A_3 &=& C_3 \,,
\label{mmet}
\eea
(here $x_{11}$ denotes the M direction, and the units are such that the M theory Planck scale is 1) which has $p_{11}$ and wrapped M5 charges.

As we proceed to smaller $r$, both the transverse $S^3$ and the $T^4$ may shrink.  The $S^3$ metric is proportional to $r^{2/3}$ in the smeared regime but constant $\rho^0$ in the unsmeared regime, the latter property following from the conformal behavior of the M5 solution.  One can check that the $S^3$ radius never falls below the coincident M5-brane value $N_5^{1/3}$, so this never leads to a breakdown of the solution.  For the $T^4$, the radii become Planckian when $H = R^2 V^{3/2} /g^2$.  In the smeared solution this is at $r_{\rm II'} = gQ^{1/2} /V^{3/4}R = r_{\rm M}/V^{3/4}$.  In the unsmeared solution it is at $\rho_{\rm II'} = g^{2/3} Q^{1/3}/V^{1/2} R$.  If $r_{\rm II'} > r_{\rm u}$ the M theory solution breaks down in the unsmeared regime at $r_{\rm II'}$, otherwise it breaks down at $\rho_{\rm II'}$.

In order to extend the solution further, we must first reduce to IIA along one of the $T^4$ directions.  The other three torus radii remain small, so a $T$-duality along these is needed next.  This leaves the IIB coupling large, so a further $S$ duality is needed.  The net result of this $STS$ transformation is a parametrically valid type II description
\bea
ds^2_{\rm II'} &=& V\left[d{x_{11}}^2+\frac{2R}{gH} dt dx_{11} +\frac{R^2}{g}\left(d\tilde y^2/R^2 +dx_4^2\right)\right]+ d \tilde z_3^2\,, \nonumber\\
e^{\Phi_{\rm II'}} &=& \frac{R V^{3/4}}{g H^{1/2}}  \,, \nonumber\\
B_2^{\rm II'} &=& 
\frac{R^2 V}{gH} dt\wedge dx_{11} \,. \label{ii'}
\eea
In this solution one of the original torus directions has become the M direction, while $x_{11}$ has emerged as a new periodic direction.  The three $\tilde z$-circles remaining from the original $T^4$ are now string-sized. We therefore label this solution simply as II$'$, since it is midway between the IIA and IIB descriptions. The charges are F-string winding in the 11-direction and $p_{11}$.

In this final form, the curvature becomes large at $r_{\rm b} = \rho_{\rm b} = g /V^{1/2} R$.  This is inside the unsmearing radius $r_{\rm u}$, so it is $\rho_{\rm b}$ that matters.  When curvature becomes large, no further string duality can save us.  However, we note that the II$'$ description and its breakdown are very similar to those of the supergravity description of the D1-brane in Ref.~\cite{Itzhaki:1998dd}.  There, the final supergravity description is in terms of F-strings, and it is argued that dynamics at smaller $r$ (lower energy) is given by the long string CFT identified in Refs.~\cite{Motl:1997th,Banks:1996my,Dijkgraaf:1997vv}.
We expect the same to hold here as well, although the additional momentum charge means that we are looking at excited states in this theory.  

This conjecture is in keeping with the general expectation that when the curvature becomes large while the string coupling goes to zero, as it does in the solution~(\ref{ii'}), one should look for a weakly coupled CFT description.  The leading twist interaction in the CFT is irrelevant~\cite{Dijkgraaf:1997vv}, so that the coupling continues to go to zero in this regime.

The full picture is summarized in Figure~\ref{fig:radii}.  Martinec and Sahakian do not restrict to the asymptotic D1-D5 regime~(\ref{regime}) and so cover a wider range of phases (Ref.~\cite{Martinec:1999sa}, Fig.~4).  Note also that they use different variables for the axes.
For such non-conformal branes~\cite{Itzhaki:1998dd}, the physics at a given scale or temperature is governed by the weakly coupled description at the corresponding holographic radius.  For example, at the lowest energies the weakly coupled field theory is the appropriate description, as it is for D$p$-branes with $p>3$.  

\subsection{Fuzzball geometries}\label{subsec:fuzzgeos}

\subsubsection{Fuzz and the onion}

A more general class of two-charge geometries is characterized by a curve $\vec{F}(v)$ in the non-compact $\R^4$~\cite{Lunin:2001jy}:
\bea
ds^2 &=& \frac{1}{\sqrt{H_1 H_5}}\left[-(dt+A)^2+(Rdy+B)^2\right]+\sqrt{H_1H_5}dx_4^2+\sqrt{\frac{H_1}{H_5}}\sqrt{V}dz_4^2
\,,\nonumber\\
e^{\Phi} &=&g\sqrt{\frac{H_1}{H_5}}
\,,\nonumber\\
C_2 &=& g^{-1} \left[ H_1^{-1} (dt+A)\wedge (R dy+B) +\zeta\right]
\,, \label{fuzzmet}
\eea
where the harmonic functions are
\bea
H_5 &=& 1 + \frac{Q_5}{L}\int_0^L \frac{dv}{|\vec{x}-\vec{F}(v)|^2}
\,,\nonumber\\
H_1 &=& 1 + \frac{Q_5}{L}\int_0^L \frac{|\dot{\vec{F}}|^2dv}{|\vec{x}-\vec{F}(v)|^2}
\,,\nonumber\\
A^i &=& \frac{Q_5}{L}\int_0^L\frac{\dot F^i dv}{|\vec{x}-\vec{F}(v)|^2} \,,  \label{fuzzharm}
\eea
with $L = \frac{2\pi Q_5}{R}$.\footnote{The range $L$ is a vestige of the original derivation of these solutions and does not have particular significance.}  The remaining quantities are defined via $dB = \star_4 dA$, $d\zeta = -\star_4 dH_5$.

To be precise, this solution describes only oscillations in the transverse directions.  The complete solution with oscillations in the torus directions is given in Refs.~\cite{Lunin:2002iz,Kanitscheider:2007wq}.  It is slightly more complicated in form, but qualitatively similar, and the same estimates of radii apply.

At $r > |\vec{F}|$ these solutions go over to the naive geometry~(\ref{naivemet}), with 
\be
Q_1 =  \frac{Q_5}{L}\int_0^L {|\dot{\vec{F}}|^2dv} \,.
\ee
Expanding $\vec{F}$ in harmonics,
\be
\vec{F} = \sum_{m=1}^\infty \vec{F}_m e^{2\pi i m v/L} + {\rm c.c.} \,, 
\ee
this becomes 
\be
2\sum_{m=1}^\infty m^2 |\vec{F}_m|^2 = \frac{Q_1 Q_5}{R^2} \label{fsum}
\ee
or
\be
\frac{2\ V R^2}{g^2} \sum_{m=1}^\infty m^2 |\vec{F}_m|^2 = N_1 N_5 .  \label{sumrule}
\ee
This last form is compatible with the quantization condition
\be
|\vec{F}_m|^2 = \frac{g^2 n_m}{2mV R^2} = \frac{r^2_{\rm b} n_m}{2m} \ \ \Rightarrow\ \ \sum_{m=1}^\infty m n_m = N_1 N_5\,,
\label{quant}
\ee
which can be derived either by duality from the F1-$p$ system~\cite{Lunin:2001jy} or by quantization of the D1-D5 solution~\cite{Rychkov:2005ji}.  Note that the breakdown radius $r_{\rm b}$ is the same as the parameter $\mu$ in the literature, meaning that $r_{\rm b}$ maps to the string length in the F1-$p$ frame.

For a solution with average harmonic $\overline{m}$, the sum~(\ref{fsum}) implies that
\be
|\vec{F}| \sim \frac{\sqrt{Q_1 Q_5}}{\overline{m}R} = \frac{\sqrt{N_1N_5}}{\overline{m}} r_{\rm b} \equiv r_{\overline{m}} \,.
\ee
As long as $r_{\overline{m}} > r_{\rm b}$ this should be a valid supergravity solution.  This translates to $\overline{m} < \sqrt{N_1N_5}$.  Note that $r_1 > r_{\rm IIA}$, so the largest solutions are described in the original IIB D1-D5 frame.  The ratio $r_1/r_{\rm nh}$ is of order $Q^{1/2}/R$.  In the asymptotic regime~(\ref{regime}) this can be either large or small, but we are usually interested in scaling up the charges with other parameters held fixed.  In this case the $\overline{m} \sim 1$ solutions extend into the flat Minkowski region.

As $\overline{m}$ decreases, the parametrically valid description of the state moves among the IIA, M, and II$'$ frames.  Since the $y$ and $z$ directions remain flat in the fuzzball solutions, it is straightforward to dualize them in the same way as for the naive solution, including the unsmearing; we do so in Appendix~\ref{app:fuzzyonion}.  For $\overline{m} > \sqrt{N_1N_5}$ the states are described by the low energy CFT rather than supergravity.  As $\overline{m} \to \infty$, the fuzzball solution approaches the naive solution, although the quantization condition puts the limit $m \leq N_1 N_5$ on the highest Fourier mode of $\vec{F}$.

For {\it typical} states, $\overline{m} \sim \sqrt{N_1N_5}$, which defines the fuzzball radius $r_{\rm f} = r_{\sqrt{N_1N_5}} = r_{\rm b}$.  
That is, these states live at the boundary of validity between the last supergravity solution and the free CFT.  The fact that these fuzzballs live at the boundary of validity of supergravity is well-known in the F1-$p$ frame~\cite{Lunin:2002qf}, and remains true here.   The duality cascade that we have found means that the D1-D5 geometries are never good descriptions of these typical fuzzball states.  The best supergravity description would be the F1-$p$ solutions~\cite{Callan:1995hn,Dabholkar:1995nc}.

Note that for both the fuzzball and naive D1-D5 geometries, the IIB curvature is always small in terms of the tension of a probe F-string, seemingly in contradiction with what we have found.
The point of the duality cascade is that there is a lighter string-like object: a probe KK monopole (charged on the $y$-circle, wrapped on the torus and extended in one transverse direction) which maps to a probe F-string in the II' picture. It has a tension $\tau_{\rm KK} \sim R_y^2(r) V(r)/g^2(r) = R^2V/g^2 H(r)$, which goes to zero as it approaches the singularity and matches the IIB curvature $Q^{-1}$ at $\rho_{\rm b}$, signaling a breakdown.

Before we go on, there is one additional radius of interest.  The two-charge system has a known microscopic entropy of order 
\be
S \sim \sqrt{N_1N_5} \,. \label{entropy}
\ee
Let us compare this to the Bekenstein-Hawking entropy that we would ascribe to a spherical shell surrounding the singularity in the naive geometry.  In the smeared regime $r R > 1$ this would be 
\be
\frac{\mbox{8d area}}{l_p^8}\sim R_{y}\times V_{S^3}\times V_{T^4}\times e^{-2\Phi} = \frac{r}{r_{\rm b} }\sqrt{N_1N_5} \,. \label{area1}
\ee
The area in Planck units is the same in any duality frame; the decomposition~(\ref{area1}) corresponds to the IIB picture.
In the smeared regime $r R < 1$ it is
\be
\frac{\mbox{8d area}}{l_p^8}\sim V_{S^4}\times V_{T^3}\times L_{11} \times e^{-2\Phi} = \frac{\rho}{\rho_{\rm b}} \sqrt{N_1N_5} \,, \label{area2}
\ee
where we have used the II$'$ description.  It is now interesting to ask, at what radius is the holographic value equal to the actual entropy? We see that this is true at $\rho = \rho_{\rm b}\equiv \rho_S$.  Again this reproduces a result known from the F1-$p$ frame~\cite{Sen:1994eb,Lunin:2002qf}, that the horizon radius corresponding to the microscopic entropy is comparable to the breakdown radius and the typical fuzzball radius.

It is not clear then whether the fuzzball solutions are any better as a description than the naive geometry.

\subsubsection{From F1-$p$ to D1-D5 and back again}
\label{subsec:f1pduals}

The D1-D5 fuzzball geometries were originally obtained~\cite{Lunin:2001jy} via U-duality from F1-$p$ geometries describing a string with left-moving excitations:
\be
\left(\begin{array}{c} $F1$\\ p\end{array}\right)\underrightarrow{S} 
\left(\begin{array}{c} $D1$\\ p\end{array}\right)\underrightarrow{T_{T^4}}
\left(\begin{array}{c} $D5$\\ p\end{array}\right)\underrightarrow{S}
\left(\begin{array}{c} $NS5$\\ p\end{array}\right)\underrightarrow{T_y T_6}
\left(\begin{array}{c} $NS5$\\ F1\end{array}\right)\underrightarrow{S}
\left(\begin{array}{c} $D5$\\ D1\end{array}\right) \,.  \label{horiz}
\ee
This relates the F1-$p$ and D1-D5 moduli as
\be
g_{\mbox{\scriptsize F1-$p$}} = \left(\frac{ V^{3/4} R}{g }\right)_{\mbox{\scriptsize D1-D5}}\,,\quad R_{\mbox{\scriptsize F1-$p$}}=\left(\sqrt{V}\right)_{\mbox{\scriptsize D1-D5}}\,,\quad V_{\mbox{\scriptsize F1-$p$}}^{1/4} = \left(\frac{\sqrt{V}}{g}\right)_{\mbox{\scriptsize D1-D5}} \,.
\ee
The F1-$p$ solutions describe the physics in a corner of the moduli space where, in terms of the asymptotic D1-D5 moduli, ${ V^{3/4} R}/{g} <1$,  ${V} > 1$, and $\sqrt{V}/g > 1$. In this regime the D1-D5 description at infinity breaks down.

It is amusing that the descent into the fuzzball core leads us back to the  F1-$p$ duality frame in which the solutions were originally obtained, a sort of ``ontogeny recapitulates phylogeny." Unlike the horizontal duality chain~(\ref{horiz}), the asymptotics are held fixed as we descend. The II$'$ frame in the deep IR is related to the asymptotic IIB frame by
\be
\left(\begin{array}{c} D5\\ D1\end{array}\right)\underrightarrow{T_y} 
\left(\begin{array}{c} D4\\ D0\end{array}\right)\underrightarrow{S_{11}}
\left(\begin{array}{c} M5 \\ p\end{array}\right)\underrightarrow{S_{6}}
\left(\begin{array}{c} D4\\ p\end{array}\right)\underrightarrow{T_{{789}}}
\left(\begin{array}{c} D1\\ p\end{array}\right)\underrightarrow{S}
\left(\begin{array}{c} F1\\ p\end{array}\right) \,,
\ee
which inverts the horizontal chain: 
$
S T_{T^4} S T_{y6} S T_{y} S_{11} S_{6} T_{789} S =1 .$

Examining the II$'$ metric, one finds $R^{11}_{\rm II'} = \sqrt{N_1/N_5} = R^y_{\mbox{\scriptsize F1-P}}$, while $R^y_{\rm II'} = g^{-1} \sqrt{N_1/N_5} = V_{\mbox{\scriptsize F1-P}}^{1/4}$. The long chain from F1-$p$ to D1-D5 and back again just switches the $(y,x^6)$ circles of the original F1-$p$ picture with the $(x_{11}, y)$ circles of II$'$: the emergent II$'$ description of D1-D5 at low energies matches the F1-$p$ description obtained by moving on the asymptotic moduli space.

\subsubsection{Orbifolds and orbifolds}

The target space of the free CFT is the orbifold $(R^4 \times T^4)^{N_5}/S_{N_5}$.  This should not be confused with the orbifold $(T^4)^{N_1 N_5}/S_{N_1 N_5}$ which also appears in the D1-D5 system.  The latter is relevant in an entirely different duality frame where $N'_5 = 1$, reached by turning on form fields on the $T^4$.  We also note some other differences between these:
\begin{itemize}
\item
For $(R^4 \times T^4)^{N_5}/S_{N_5}$ we are interested in states with $N_1$ left-moving excitations.  For $(T^4)^{N_1 N_5}/S_{N_1 N_5}$ we are interested in ground states.
\item
For $(R^4 \times T^4)^{N_5}/S_{N_5}$ we are only interested in the sector with a single long string, because only this corresponds to a single-particle state.  For $(T^4)^{N_1 N_5}/S_{N_1 N_5}$ the fractionalized strings are all bound to the D5-branes, so all winding sectors correspond to single-particle states.
\item
For $(R^4 \times T^4)^{N_5}/S_{N_5}$ the twist interaction is irrelevant as noted above.  For $(T^4)^{N_1 N_5}/S_{N_1 N_5}$ it is marginal.
\end{itemize}

\subsubsection{Lessons}\label{subsubsec:lessons}

Our conclusion is that the typical fuzzball is at the transition between two descriptions, a supergravity description with stringy sources and a weakly coupled CFT description.   There is yet a third description that has been given for this system: the black hole solution with a horizon, which exists when higher derivative terms are included~\cite{Dabholkar:2004yr,Dabholkar:2004dq}.  This is usually discussed in systems with half as much supersymmetry, where the $T^4$ is replaced by K3, but as shown in Appendix~\ref{app:K3} the onion structure is the same in this case.\footnote{We thank Nori Iizuka for discussions of the K3 case and the relation between different pictures.}  This solution allows a precise counting of supersymmetric states, but like the naive and fuzzball geometries it is on the boundary of its range of validity.  

We are primarily interested in regimes where the fuzzball geometries are parametrically valid, and we will find one in \S~\ref{sec:Jneq0}, but here we make a few remarks about the marginal case found above.  Ref.~\cite{Sen:2009bm} argues that two-charge systems fall into two classes, those whose description is given by smooth horizonless solutions, and those where it is a black hole from a higher derivative action.  The D1-D5 system was argued to be of the first type, but the onion structure shows that, if this classification is correct, then it is of the second type.

The fuzzball description might seem to retain more information by distinguishing individual microstates, but this information may not be meaningful.  As argued in~\cite{Sen:2009bm}, interactions mix the BPS states of interest into a larger space of non-BPS states, so that the resulting BPS states may bear little resemblance to their naive form.  This phenomenon can be seen for example in the low-energy CFT frame.  There is a twist interaction, which mixes the BPS single-long-string sector with non-BPS multi-string states (these are somewhat localized in the transverse directions and so have supersymmetry-breaking $p_\bot$).

However, there is an interesting counterargument.  The one-point functions of chiral operators distinguish microstates~\cite{Skenderis:2006ah,Kanitscheider:2006zf}, and these one-point functions are not renormalized~\cite{Baggio:2012rr}.\footnote{In Ref.~\cite{Giusto:2014aba}, it has been shown that these same one-point functions imply that the entanglement entropy distinguishes microstates.}
It is puzzling to reconcile this with the point of view above.  Note that in a Haar-random state the one-point functions will be of order $e^{-S/2}$~\cite{Balasubramanian:2007qv}.  Curiously, the same is true for Schwarzschild black holes.  In thermal systems, variations of the one-point functions from their thermal averages are of order $e^{-S/2}$~\cite{shredder}.   However, this implies that the eigenvalues are $O(1)$, and one can find a basis in which the one-point functions are of this size in {\it any} thermal system.

Indeed,  a similar basis has been used to argue for the genericity of firewalls, namely the basis in which the Hawking occupation numbers are diagonal~\cite{Bousso:2013wia,Marolf:2013dba}.  These would be analogous to number eigenstates for the $\vec F_m$.   So the `firewall' basis in these papers seems to be the Schwarzschild equivalent of the two-charge fuzzball states.  This parallel is somewhat unexpected, since extremal and non-extremal horizons are in many respects quite different.  Clearly it is interesting to contemplate this further.

\section{The $J > 0$ system}\label{sec:Jneq0}

\subsection{Naive geometry: small black ring}

We now focus on fuzzball states having angular momentum $J$ in the 1-2 plane of the transverse space.  The maximum value $J_{\rm max} = N_1 N_5$ corresponds to the classical solution~\cite{Maldacena:2000dr,Lunin:2001fv}
\be
\vec{F}_{\rm max}=(a \cos \omega v,a \sin \omega v)\,,
\ee
where only the $m=1$ harmonic is excited.  Here
\be
a = r_1 = \sqrt{Q_1 Q_5}/{R}\,,\quad \omega = 2\pi/L = R/Q_5 \,.  
\ee

For near maximal $J$, i.e.\ 
\be
\epsilon \equiv \frac{J_{\rm max} - J}{J_{\rm max}} \ll 1\,,
\ee
 most of the excitation goes into the first harmonic.  Such a solution can be described by the profile
\bea
 \vec{F} &=&  \vec{F}^{(0)} + \delta \vec{F} \,, \nonumber\\
 \vec{F}^{(0)} &=& (a_0\cos \omega v,a_0 \sin \omega v)\,,\label{ringprofile}
\eea
with $a_0= a \sqrt{J/J_{\rm max}}$.  The sum rule~(\ref{sumrule}) gives
\be
\frac{2\ V R^2}{g^2} \sum_{m=1}^\infty m^2 |\delta  \vec{F}_m|^2 = \epsilon N_1 N_5 \,. \label{ringsum}
\ee
For typical states, the dominant harmonic is then $\overline{m} \sim \sqrt{\epsilon N_1 N_5 }$.  We have $|\delta \vec{F}|/|\vec{F}^{(0)}| \sim \sqrt{\epsilon}/\overline{m} \sim 1/\sqrt{N_1 N_5}$, so the geometry is a fuzzy ring, with thickness much less than its radius.

As in the $J=0$ case, we can think of the naive geometry as obtained by taking the $\overline{m} \to \infty$ limit, or equivalently by interpolating the geometry outside the fuzz down to the core of the ring.  This gives~\cite{Lunin:2002qf,Balasubramanian:2005qu}
\bea
H_5 &\approx& 1 + \frac{Q_5}{L}\int_0^L \frac{dv}{|\vec{x}-\vec{F}^{(0)}(v)|^2}
\,,\nonumber\\
H_1 &\approx& 1 + \frac{J_{\rm max}}{J}\frac{Q_5}{L}\int_0^L \frac{|\dot {\vec{F}}^{(0)}|^2dv}{|\vec{x}-\vec{F}^{(0)}(v)|^2}
\,,\nonumber\\
A^i &\approx& \frac{Q_5}{L}\int_0^L\frac{\dot F^{(0)i} dv}{|\vec{x}-\vec{F}^{(0)}(v)|^2} \,,
\eea
which is shown in~\cite{Iizuka:2005uv,Balasubramanian:2005qu} to be a special case of the black ring~\cite{Elvang:2004rt,Bena:2004de,Gauntlett:2004qy}.  

Because of the factor of ${J_{\rm max}}/{J}$, the cancellation of singular behaviors that gives rise to a smooth geometry~\cite{Maldacena:2000dr,Lunin:2002iz}  no longer occurs, and there is a singularity in the core of the ring. Using ``ring coordinates'' as in~\cite{Elvang:2004rt,Bena:2004de} the flat metric $dx_4^2$ on $\mathbb{R}^4$ is
\be
dx_4^2=\frac{a_0^2}{(X-Y)^2} \left[\frac{dY^2}{Y^2-1} +(Y^2-1) d\psi^2 + \frac{dX^2}{1-X^2} +(1-X^2) d\phi^2\right]\,, \label{fullsol}
\ee
and $\mathbb{R}^4$ is foliated by surfaces of constant $Y$ with topology $S^1 \times S^2$. The coordinates $X,Y$ take values in the range $-1 \leq X \leq 1$ and $-\infty < Y \leq -1$ and $\psi,\phi$ are polar angles in two orthogonal planes in $\mathbb{R}^4$ with period $2\pi$. The angle $\psi$ is along the ring and the ring singularity is located at $Y=-\infty$.
In terms of the ring coordinates we have
\be
 H_{1}=1+\frac{Q_{1}}{\Sigma}\,, \quad H_{5}=1+\frac{Q_{5}}{\Sigma}\,, \quad \text{where} \quad \Sigma=\frac{2a_0^2}{X-Y}\,,
\ee
and
\be
A_\psi = \frac{R}{2} (1+Y)\,, \quad B_\phi= \frac{R}{2} (1+X)\,,\quad  \zeta_{\psi \phi} = \frac{Q_5}{2}\left[Y-\frac{1-Y^2}{X-Y}\right]\,.
\ee

In the near-ring limit it is useful to switch from the ring coordinates $X,Y$ to $\theta,x_\bot$ :
\be
X\approx -\cos\theta\,, \qquad 1+Y \approx -\frac{a_0}{x_\bot}\,, 
\ee
where the angle coordinate $\theta$ combines with $\phi$ to form an $S^2$ and $x_\bot$ is the radial coordinate transverse to the ring. The ring singularity is now located at $x_\bot=0$. The leading behaviors (simplified again to $Q_1 = Q_5 = Q$) are
\be
H_5 = H_1 \approx \frac{R}{2cx_\bot}  \,,\quad
A_\psi \approx -\frac{Q c}{2x_\bot }\,,\quad
B_\phi \approx \frac{R}{2}(1-\cos\theta) \,,\quad
\zeta_{\psi \phi} \approx - \frac{Q}{2}(1-\cos\theta)\,,
\ee
where we have introduced $c = \sqrt{J/J_{\rm max}}= \sqrt{1-\epsilon}$.
The naive near-ring metric becomes 
\begin{eqnarray}
ds_{\rm near}^2 &\approx & \frac{2 c x_\bot}{R} \Big[-\Big(dt-\frac{Qc}{2 x_\bot} d\psi \Big)^2 + R^2 \left(d{y}+\frac{1-\cos\theta}{2}  d\phi\right)^2\Big]  \nonumber\\
&&+ \frac{Rc}{2x_\bot} \Big[ dx_\bot^2 + x_\bot^2 (d\theta^2 +\sin^2\theta d\phi^2) \Big]+
\frac{cQ^2}{2Rx_\bot} d\psi^2 +  \sqrt{V} dz_4^2\,\,.\label{nearring}
\end{eqnarray}
For $c=1$ this is smooth at $x_\bot = 0$, but for $c < 1$ it becomes singular there. The near-ring dilaton is simply $e^{\Phi} = g$ and the RR potential is given by
\be
C_2 \approx 2c x_\bot dt \wedge \left[dy + \frac{1-\cos\theta}{2} d\phi\right] + Q c^2 \left[ dy + \left(1+\frac{1}{c^2}\right) \frac{1-\cos\theta}{2} d\phi\right]\wedge d\psi\,.
\ee
In the near-ring limit there are four local charges corresponding to D1 and D5 branes wrapped on the $y$ circle and the torus, KK monopoles wrapping the $y\psi$ directions and the torus and momentum charge along the $\psi$ direction.\footnote{Note that in the near-ring geometry~(\ref{nearring}) the circumference of the $\psi$-circle seems to go to zero at large~$x_\bot$.  However, this occurs outside of the range of validity of~(\ref{nearring}).  In the full solution~(\ref{fullsol}) the 1's in the harmonic functions prevent the $\psi$-circle from shrinking.}

\subsection{No black onion rings}\label{subsec:noonionring}

As we proceed toward smaller $x_\bot$, the $y$-circle again shrinks.  However, this is merely a coordinate effect: the metric in the $x_\perp$-$y$ plane is just $\R^2$, with $y$ an angular coordinate.  A $T$-duality provides a useful description only if the shrinking circle does not cap off smoothly,
as in the $J=0$ metric~(\ref{naivemet}).  Hence there is no repetition of the layered structure found before: there is no black onion ring.

The first breakdown of the naive geometry~(\ref{nearring}) is due to the divergence of the curvature, because of the uncanceled $1/x_\bot$ in $g_{\psi\psi}$ and the squashing of the Hopf fibration.  The curvature invariant is calculated to be
\begin{equation}
R_{\mu\nu\rho\sigma} R^{\mu\nu\rho\sigma} = \frac{22}{R^2 x_\bot^2} \epsilon^2\,.
\end{equation}
This defines the breakdown radius~$x_{\bot\rm b} = \epsilon/R$.

As for the $J=0$ case there are two other radii to compare.  From the discussion below Eq.~(\ref{ringsum}) it follows that the fuzzball radius is 
\be
x_{\bot\rm f} \sim r_1/\sqrt{N_1 N_5} = g/R\sqrt V \,.
\ee
  To obtain the entropy radius, the area in Planck units of a torus surrounding the ring is
\be
\frac{\mbox{8d area}}{l_p^8}\sim L_{\psi}\times L_{y} \times L_{S^2}\times L_{T^4} \times e^{-2\Phi} \sim  Q \sqrt V x_\bot R \sqrt{\epsilon}/g^2
\,. \label{arearing}
\ee
Equating this to the entropy $\sqrt{\epsilon N_1 N_5}$, we obtain $x_{\bot S} = x_{\bot\rm f} = g/R\sqrt V$.

The matching of the fuzzball and entropy radii for the ring has been noted previously~\cite{Lunin:2002qf}.  But unlike the $J=0$ case considered above, the breakdown radius differs from these:
\be
\frac{x_{\bot \rm b}}{x_{\bot {\rm f}}} = \frac{x_{\bot \rm b}}{x_{\bot S}} = \frac{\epsilon \sqrt{V}}{g} \,.
\ee
This ratio can be either large or small.

The interesting case is when ${x_{\bot {\rm f}, S} \gg x_{\bot \rm b}}$: the fuzzballs appear at a radius where the curvature is still small.\footnote{
The curvature is smaller than the $1/\mu^2$ that might have been expected from the curvature in the original F1-$p$ frame ($\mu$ is defined below Eq.~(\ref{quant})).  This happens because terms arising originally from $B_{\mu\nu}$ combine with the metric to produce a smoother Hopf-fibered metric.  In the parameter regime where the F1-$p$ duality frame applies, the curvature becomes stringy and there is a higher-derivative black hole solution~\cite{Dabholkar:2006za}. }  
Thus they are good supergravity solutions, and give a parametrically valid description of the states in this regime.  It is interesting to ask whether the naive geometry shows any signs of this premature breakdown.  

For comparison, in the enhan\c{c}on~\cite{Johnson:1999qt}  and the ${\cal N}=1^*$ geometries~\cite{Polchinski:2000uf}, singularities are resolved by branes expanding out to radii where the naive curvature is small.  In these cases, brane probes give an indication of this: if one tries to add branes to the singularity, they feel a repulsive potential at radii where the curvature is still small.  This does not seem to be the case for the black ring: one can consider atypical solutions with larger harmonics, and these can approach the ring much more closely.  In the Klebanov-Tseytlin geometry~\cite{Klebanov:2000nc}, resolved in supergravity~\cite{Klebanov:2000hb}, a flux takes an unphysical negative value at finite radius; nothing analogous happens here.

The signal of the breakdown of the naive geometry for the black ring seems to be the entropy radius.  If the naive geometry were valid, we could consider a torus thinner than $x_{\bot S}$, and the number of quantum states contained within would be larger than the exponential of the Bekenstein-Hawking entropy for the torus.  It is natural to conjecture that this cannot happen: that if a system has a Hilbert space of dimension~${\cal D}$, then the states must be distinguishable at a radius where a surrounding surface has area $\log \cal D$, in Planck units. 

For ${x_{\bot {\rm f}, S} \ll x_{\bot \rm b}}$, we have not yet found a good description.

\section{Discussion}\label{sec:discussion}

Our study of two-charge fuzzballs has led to some surprises.  

For $J=0$, we find that the appropriate duality frame depends on the size of the fuzzball state, which is determined by the average harmonic $\overline{m}$.  For typical states, the best supergravity description is not in terms of smooth D1-D5 solutions but rather has stringy sources.  We emphasize the importance of three radii: the radius of the typical fuzzball, the radius where the transverse area is equal to the microscopic entropy, and the radius where the curvature approaches the string scale. For the two-charge system, these three radii agree, meaning in particular that the supergravity description is beginning to break down for typical states. This triple agreement is well-known in the original F1-$p$ duality frame; it is therefore unsurprising to find it here since the II$'$ frame with F1-$p$ charges is actually the correct duality frame for the typical fuzzball.

Fuzzballs with other values of $\overline{m}$ are parametrically valid in one of the supergravity pictures, or in the free \mbox{CFT}. These descriptions accurately capture dynamical behavior and excited states, not just BPS properties.

For three-charge black holes the entropy is $S_{\mbox{\scriptsize 3-charge}}  \sim \sqrt{N_p N_1 N_5}$.  When $N_p \ll N_1, N_5$, the geometry resembles the two-charge geometry at large radius.  It begins to differ at the entropy radius~(\ref{area1}, \ref{area2}) that would correspond to $S_{\mbox{\scriptsize 3-charge}}$.  This is 
\be
r_{\mbox{\scriptsize 3-charge}}(N_p) \sim \sqrt{N_p}\, r_{\rm b} \,.
\ee
We see that the correct description of these solutions can be any of IIB, IIA, M, or II$'$, depending on $N_p$.

For $J \neq 0$, we have found a regime near $J_{\rm max}$ where the fuzzball solutions are of low curvature.  It is interesting that the naive solution gives no direct indication of breakdown at the corresponding radius.  The curvature is small, and probe branes see no breakdown.  The key indicator seems to be the entropy radius: if the naive geometry were the correct description down to smaller radii, there would not be room for all the microstates. This leads us to conjecture that if some sets of microstates give rise to a common geometry, then this geometry must break down when the transverse area is of order the entropy in Planck units.

If we apply this to the Schwarzschild geometry in a naive way, the entropy radius $r_{\rm S}$ is the Schwarzschild radius $r_{\rm s}$.  If we pass through this radius into the interior where $r < r_{\rm s}$, there are then too many microstates unless we begin to see deviations from the Schwarzschild geometry: this is the fuzzball proposal.  Of course it is a speculation to extend such a principle from the two-charge geometry to Schwarzschild, but we have noted other parallels in \S~\ref{subsec:fuzzgeos}.\\

{\bf Acknowledgments}\\
We would like to thank J.~Maldacena, E.~Martinec, and S.~Mathur for helpful discussions. We have also benefited from discussions with N.~Iizuka, K.~Skenderis, M.~Taylor and other participants at the Aspen Center for Physics, supported in part by National Science Foundation Grant No. PHYS-1066293.
F.C. is supported by National Science Foundation Grant No. PHY11-25915.  B.M. is supported by the NSF Graduate Research Fellowship Grant No. DGE-1144085.  J.P. is supported by National Science Foundation Grant Nos.  PHY11-25915 and PHY13-16748. A.P. is supported by National Science Foundation Grant No. PHY12-05500.

\appendix

\section{Black onions on K3}\label{app:K3}

Taking the D1-D5 system to live on K3 instead of a $T^4$, we find a heterotic theory at the core of the onion.\footnote{We thank Nori Iizuka for asking about this case.}
This is as expected, since the duality chain of \S~\ref{subsec:f1pduals} sending type II F1-$p$ to D1-D5 on $T^4$ maps heterotic F1-$p$ to D1-D5 on K3.

Starting from the naive metric~(\ref{naivemet}) with K3 replacing the torus, one is led along the same duality chain until the K3 becomes small in the M theory description. Past this point, string-string duality suggests that the appropriate picture is the heterotic theory on $T^3$.  This follows from the same $STS$ series that we used before, but now the duals go through a IIA orientifold, type I, and then heterotic $SO(32)$~\cite{Sen:1996yy}.  The transformations on the metric, $B$-field, and dilaton are the same as before, so 
we obtain
\bea
ds^2_{\rm het}&=&V\left[d{x_{11}}^2+\frac{2R}{gH}dt  dx_{11} +\frac{R^2}{g}\left(d \tilde y^2/R^2 +dx_4^2\right)\right] + d\tilde{z}_3^2
\,,\nonumber\\
e^{\Phi_{\rm het}} &=& \frac{R V^{3/4}}{g H^{1/2}} \,,\nonumber\\
B_2^{\rm het} &=& \frac{R^2 V}{gH} dt\wedge dx_{11} \,. 
\eea
This matches the II$'$ solution (\ref{ii'}) exactly; the only difference from the $T^4$ case is that we have ended up in a heterotic theory. As before, this description is parametrically valid until $\rho_{\rm b}$, where the curvature becomes large.

\section{Fuzzy onions}\label{app:fuzzyonion}

We repeat the analysis of \S~\ref{subsec:naivegeo} for the fuzzball geometries, obtaining descriptions valid for fuzzballs with various values of $m$.

Starting from the IIB frame with fuzz (\ref{fuzzmet}),

\bea
ds_{\rm IIB}^2 &=& H^{-1}\left[-(dt+A)^2+(Rdy+B)^2\right]+H dx_4^2+\sqrt{V}dz_4^2
\,,\nonumber\\
e^{\Phi_{\rm IIB}} &=&g
\,,\nonumber\\
C_2 &=& g^{-1} \left[ H^{-1} (dt+A)\wedge (R dy+B) +\zeta\right]
\,,
\eea
the IIA fuzzball geometry is

\bea
ds_{\rm IIA}^2 &=& -H^{-1} (dt+A)^2 + H\left[d\tilde y^2/R^2 +dx_4^2\right]+\sqrt{V}dz_4^2
\,,\nonumber\\
e^{\Phi_{\rm IIA}} &=&g \sqrt{H}/R
\,,\nonumber\\
B^{\rm IIA}_2 &=& R^{-1} B\wedge d\tilde y\,,\nonumber\\
C_1 &=& \frac{R}{gH} (dt+A)
\,,\nonumber\\
C_3&=& g^{-1}\zeta \wedge d\tilde y
\,.
\eea
The B-field corresponds to NS5 dipole charge along $\vec{F}$, $T$-dual to the KK dipole in IIB. The branes unsmear for $r<r_u$ just as in the naive geometry.

The fuzzy IIA becomes strongly coupled beyond $r_M/\rho_M$, suggesting an M theory description:
\bea
ds^2_{\rm M} &=& e^{-2\Phi_{\rm IIA}/3} ds^2_{IIA} + e^{4\Phi_{\rm IIA}/3}(dx_{11}+ C_1)^2
\nonumber\\
&=& \left(\frac{R^2}{g^2 H}\right)^{1/3}\left\{-H^{-1} (dt+A)^2 + H\left[d\tilde y^2/R^2 +dx_4^2\right]+\sqrt{V}dz_4^2\right\}
\nonumber\\
&&+ \left(\frac{g^2 H}{R^2}\right)^{2/3}\left[ dx_{11} + \frac{R}{gH}(dt+A)\right]^2
\,,\nonumber\\
A_3 &=& C_3 + B^{\rm IIA}_2 \wedge dx_{11}\,,
\eea
with the NS5 lifting to M5 dipole.

Once again the torus becomes small past $r_{II'}/\rho_{II'}$, and performing an $STS$ transformation as for the naive geometry yields fuzzy II$'$:
\bea
ds^2_{\rm II'}&=&V\left[d{x_{11}}^2+\frac{2R}{gH}(dt+A) dx_{11} +\frac{R^2}{g}\left(d\tilde y^2/R^2 +dx_4^2\right)\right]+ d\tilde{z}_3^2
\,,\nonumber\\
e^{\Phi_{\rm II'}} &=& \frac{R V^{3/4}}{g H^{1/2}}\,,
\eea
and $B_2^{\rm II'}$ whose field strength satisfies $H_3^{\rm II'} =\star d\left(A_3\wedge d\tilde z_3\right)$.
The M5 dipole descends to F1 dipole in the final frame, localized along $\vec{F}$.

\end{document}